\documentclass[10pt,twocolumn,superscriptaddress,prl]{revtex4}
\usepackage{nicefrac}       %For fractions that aren't ugly.  See also package "units"
\usepackage[tight,nice]{units}
\usepackage{graphicx}       %For Graphics support
\usepackage{amsmath,amsfonts,amssymb}
\usepackage{mathcomp}   %mainly to get a non-ugly \tcmu for microns

\newcommand{\fh}{\ensuremath{\nu=\nicefrac{5}{2}}}
\newcommand{\fhf}{\ensuremath{\nicefrac{5}{2}}}
\newcommand{\dipR}{\ensuremath{2\nicefrac{1}{3}}}
\newcommand{\dipL}{\ensuremath{2\nicefrac{2}{3}}}

\newcommand{\um}{\ensuremath{\tcmu\mathrm{m}}}

\newcommand{\Rd}{\ensuremath{R_\mathrm{D}}}
\newcommand{\Rl}{\ensuremath{R_\mathrm{L}}}
\newcommand{\Rxx}{\ensuremath{R_\mathrm{xx}}}
\newcommand{\Rxy}{\ensuremath{R_\mathrm{xy}}}
\newcommand{\Nbulk}{\ensuremath{N_\mathrm{bulk}}}
\newcommand{\Nqpc}{\ensuremath{N_\mathrm{QPC}}}

\newcommand{\twodeg}{2DEG} %{2\textsc{deg}}
\newcommand{\runit}{\ensuremath{h/e^2}}
\newcommand{\Vg}{\ensuremath{V_{\mathrm{g}}}}
\newcommand{\nubulk}{\ensuremath{\nu_{\mathrm{bulk}}}}
\newcommand{\nuqpc}{\ensuremath{\nu_{\mathrm{QPC}}}}
\newcommand{\qpc}{{\sc qpc}}
\newcommand{\Iac}{\ensuremath{I_{\mathrm{ac}}}}
\newcommand{\Idc}{\ensuremath{I_{\mathrm{dc}}}}

\newcommand{\fqhe}{{\sc fqhe}}
\newcommand{\iqhe}{{\sc iqhe}}
\newcommand{\sem}{{\sc sem}}
\begin{document}
\title{Fractional quantum Hall effect in a quantum point contact\\ at filling fraction \fhf}
\author{J. B. Miller}
\affiliation{Division of Engineering and Applied Science, Harvard
Univerisity, Cambridge, Massachusetts 02138}%%
\author{I. P. Radu}
\affiliation{Department of Physics, Massachusetts Institute of
Technology, Cambridge, Massachusetts 02139}%%
\author{D. M. Zumb\"uhl}%%
\affiliation{Department of Physics, Massachusetts Institute of
Technology, Cambridge, Massachusetts 02139}%%
\affiliation{Department of Physics and Astronomy, University of
Basel, Klingelbergstrasse 82, CH-4056 Basel, Switzerland}
\author{E. M. Levenson-Falk}%%
\affiliation{Department of Physics, Harvard University, Cambridge,
Massachusetts 02138}
\author{M. A.  Kastner}%%
\affiliation{Department of Physics, Massachusetts Institute of
Technology, Cambridge, Massachusetts 02139}%%
\author{C. M. Marcus}%%
\affiliation{Department of Physics, Harvard University, Cambridge,
Massachusetts 02138}
\author{L. N. Pfeiffer}%%
\author{K. W. West}
\affiliation{Bell Labs, Lucent Technologies, Murray Hill, New Jersey
07974}
\date{\today}

\begin{abstract}
Recent theories suggest that the excitations of certain quantum Hall
states may have exotic braiding statistics which could be used to
build topological quantum gates.  This has prompted an experimental
push to study such states using confined geometries where the
statistics can be tested. We study the transport properties of
quantum point contacts (\qpc s) fabricated on a GaAs/AlGaAs two
dimensional electron gas that exhibits well-developed fractional
quantum Hall effect, including at bulk filling fraction
$\nubulk=\fhf$.  We find that a plateau at effective filling factor
$\nuqpc=\fhf$ is identifiable in point contacts with lithographic
widths of \unit[1.2]{\um} and \unit[0.8]{\um}, but not
\unit[0.5]{\um}.  We study the temperature and dc-current-bias
dependence of the $\nuqpc= \fhf$ plateau as well as neighboring
fractional and integer plateaus in the \qpc\ while keeping the bulk
at $\nubulk=3$. Transport near $\nuqpc=\fhf$ in the \qpc s is
consistent with a picture of chiral Luttinger liquid edge-states
with inter-edge tunneling, suggesting that an incompressible state
at $\nuqpc=\fhf$ forms in this confined geometry.
\end{abstract}

\maketitle

\noindent The discovery \cite{Willett1987} of a fractional quantum
Hall effect (\fqhe) at the even-denominator filling fraction \fh\
has sparked a series of experimental
\cite{eisenstein1988,eisenstein1990,Pan1999Exact,pan1999Tilt,lilly1999}
and theoretical \cite{haldane1988,morf1998,Rezayi2000} studies,
leading to a prevailing interpretation of the \fhf\ state as
comprised of paired fermions condensed into a BCS-like state
\cite{MooreRead,greiter1991,ReadGreen2000,scarola2000}. Within this
picture, excitations of the \fhf\ ground state possess nonabelian
statistics \cite{NayakWilczek1996,Simon2003,stern:205338} and
associated topological properties. The possibility that such a
topological state can be accessed in the laboratory has prompted
recent theoretical work aimed at experimentally testing the
nonabelian character of the \fhf\ state
\cite{stern:016802,bonderson2006,hou:146802,chung:245311,feldman2006},
and building topologically protected quantum gates controlled by
manipulating the excitations of the \fhf\ state
\cite{Kitaev2006,bonesteel2005,dassarma2005}.

While proposed tests of the statistics of excitations of the \fhf\
state make use of confined ($\sim$ few micron) geometries, previous
studies of the \fhf\ state have been conducted in macroscopic
(\unit[100]{\um} - \unit[5]{mm}) samples. Although experiments using
mesoscopic samples with a quantum point contact (\qpc) are now
routine, the \fhf\ state is exceptionally fragile; only the highest
quality GaAs/AlGaAs heterostructures exhibit a \fhf\ state even in
bulk samples. Experimental investigation of the statistics of the
\fhf\ ground state is crucial, especially since alternative models
have been proposed to explain the \fhf\ state in confined geometries
\cite{harju2006} and in the bulk \cite{ReadGreen2000,toke2006}.

\begin{figure}[b]
\center
\includegraphics{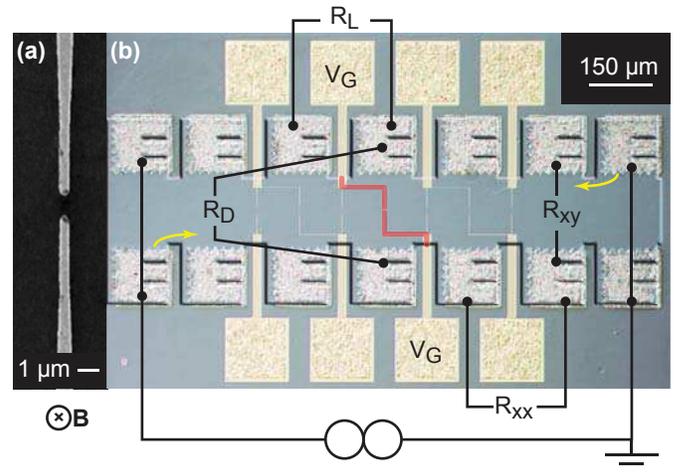}
\caption{{\label{Device}} Device and measurement setup. (a) \sem\
micrograph of the \unit[0.5]{\um} \qpc. (b) Optical micrograph of
the entire device (the outline of the wet-etched Hall bar has been
enhanced for clarity). The measurement circuit for the
red-highlighted \qpc\ is drawn schematically, with the direction of
the edge-current flow indicated by the yellow arrows.}
\end{figure}

In this paper we study the \fhf\ state in the vicinity of a quantum
point contact. Near a \qpc, the electron density is not uniform, so
the notion of a \qpc-filling fraction is not well defined. However,
based on transport measurements, it is possible to define an
effective filling fraction in the vicinity of the \qpc\ (\nuqpc), as
discussed below. Below \unit[30]{mK}, a plateau-like feature with
diagonal resistance (also defined below) near, but above, the bulk
quantized value of \unit[0.4]{\runit} is evident at $\nuqpc=\fhf$ in
\qpc s with \unit[1.2]{\um} and \unit[0.8]{\um} spacings between the
gates. On this plateau, we find a peak in the differential
resistance at dc-current bias $\Idc=0$ and a dip around
$\Idc\sim\unit[1.2]{nA}$, a characteristic shape that is consistent
with \qpc -induced quasiparticle tunneling between fractional edge
states \cite{FendleyPRB1995}.  We also observe a zero-bias peak at
$\nuqpc=\dipR$, whereas we find a zero-bias dip near $\nuqpc=\dipL$
(not shown), consistent with previous \qpc\ studies for $\nuqpc<1$
\cite{roddaro:156804}. As the temperature increases from
\unit[30]{mK} to \unit[70]{mK}, the plateaus in the \qpc\ disappear.
Fractional plateaus are not observed in a \unit[0.5]{\um} \qpc, and
the \Idc\ characteristic is flat for all magnetic fields. Together,
these observations suggest that the \fhf\ state is destroyed in the
\unit[0.5]{\um} \qpc, but can survive and exhibit quasiparticle
tunneling \cite{fendley2006,roddaro:156804,d'agosta:086801} in the
larger \qpc s.

\begin{figure}[t]
\center
\includegraphics{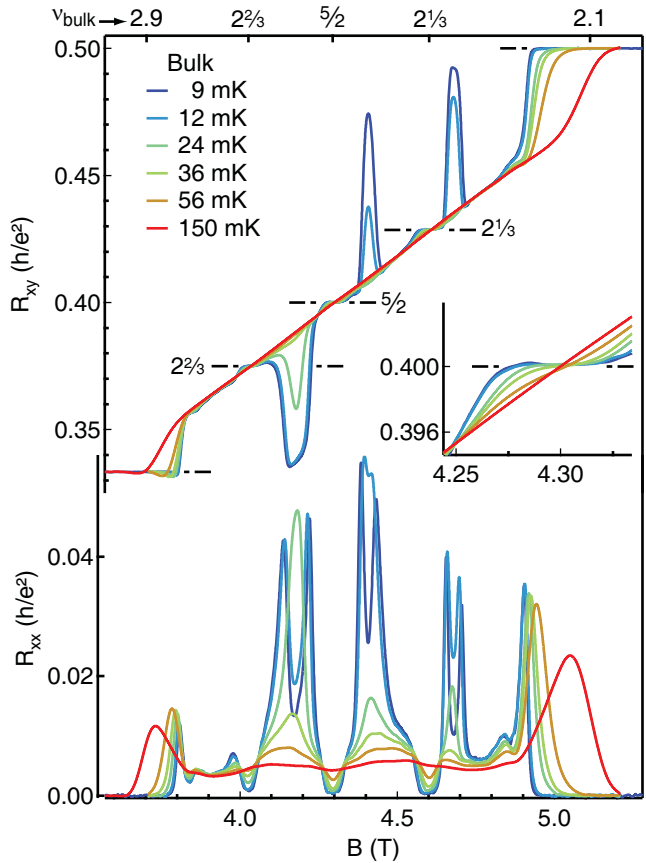}
\caption{{\label{Bulk}} Bulk transport measurements, including $T$
dependence ($T$ refers to the temperature of the refrigerator). The
inset is an enlargement of the \Rxy\ data near $\nubulk=\fhf$. }
\end{figure}

\begin{figure*}[t]
\center
\includegraphics{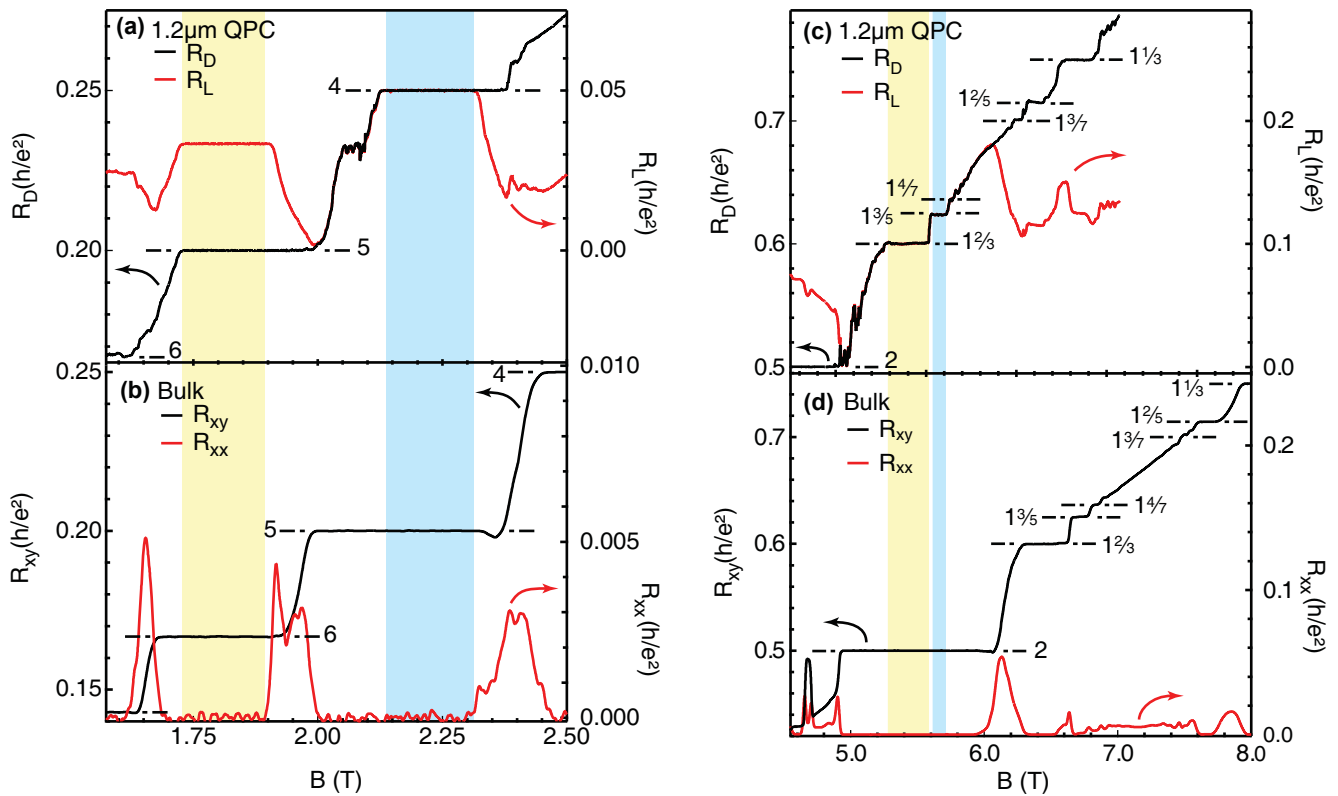}
\caption{{\label{LowB}} Typical magnetoresistance curves measured
concurrently in the \qpc\ and the bulk at low magnetic field (a and
b) and high magnetic field (c and d). Quantized resistance values
are indicated in units of \runit. The colored stripes indicate field
ranges where one quantum Hall state exists in the bulk while a
different quantum Hall state exists in the bulk. All data is for
$T\sim\unit[8]{mK}$.}
\end{figure*}

\Rxy, \Rxx, \Rd\ and \Rl\ (Fig.~\ref{Device}) are four-wire
differential resistances ($R=dV/{dI_{\mathrm{ac}}}$), measured at
$\Idc=0$ unless otherwise noted. In the \iqhe\ regime, these
resistances can be readily interpreted in terms of edge channels
\cite{Buttiker1986,BandVRd}, where $\Nbulk$ is the number of edge
channels in the bulk and $\Nqpc$ ($\leq \Nbulk$) is the number
traversing the \qpc. The bulk Hall resistance, $\Rxy \sim \runit
(1/\Nbulk)$, probes the number of edge states in the bulk region. In
the absence of tunneling across the Hall bar $\Rxy=h/e^2(1/\Nbulk)$.
The bulk longitudinal resistance, \Rxx, vanishes when \Rxy\ shows a
plateau. The diagonal resistance across a \qpc,
$\Rd\sim\runit(1/\Nqpc)$, is sensitive only to the number of edge
channels traversing the \qpc, and hence provides a \qpc-analog to
the bulk \Rxy. The longitudinal resistance across the \qpc,
$\Rl\sim\Rd-\Rxy$, contains information about both the bulk and the
\qpc-region, and is not directly analogous to the bulk \Rxx. On bulk
\iqhe\ plateaus, the filling fraction is equivalent to the number of
edge states, $\nubulk=\Nbulk$. By analogy, in the \qpc, where the
filling fraction is not well defined due to nonuniform density, we
define an effective filling fraction in the \qpc:
$\nuqpc\sim\runit(1/\Rd)$.

\begin{figure}[t]
\center
\includegraphics{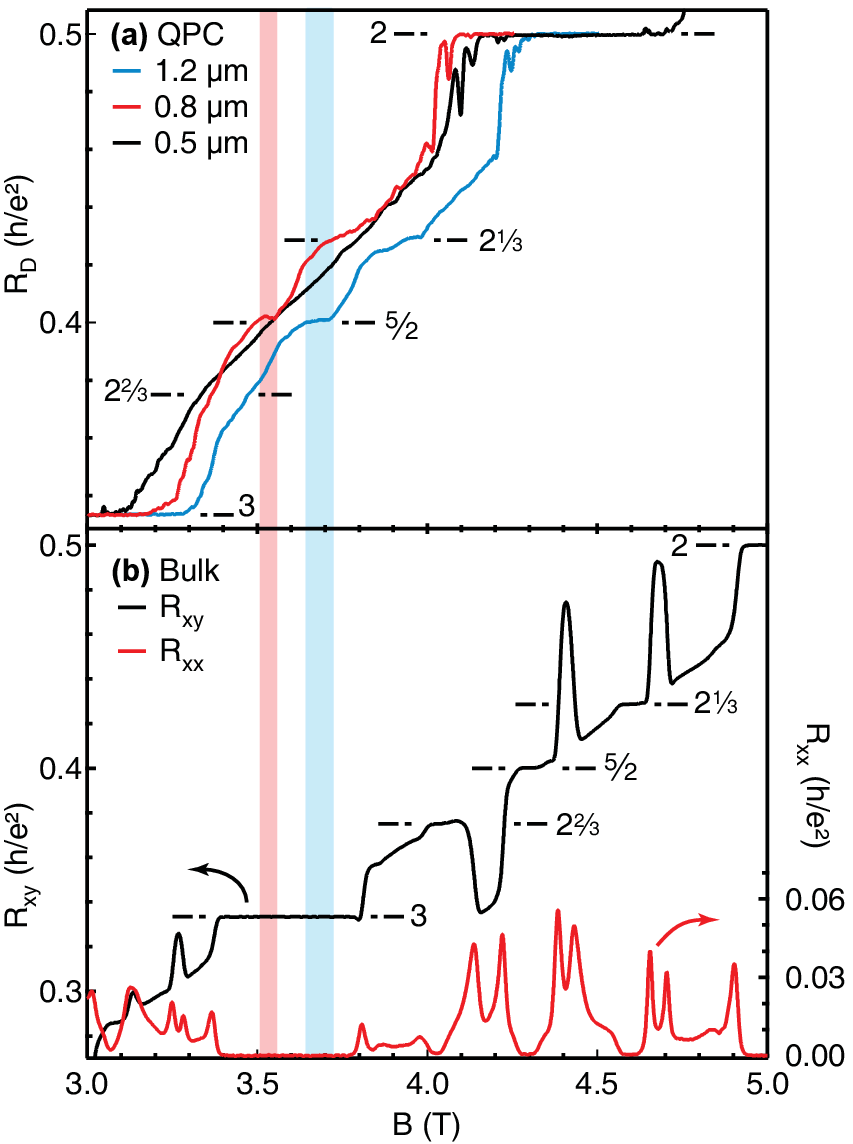}
\caption{{\label{ManyQPC}} Typical magnetoresistance from $\nu=3$ to
$\nu=2$, measured concurrently in the \qpc\ (a) and the bulk (b). In
(a), the \Rd\ curves are from three different \qpc s, of
lithographic size \unit[0.5]{\um} (black), \unit[0.8]{\um} (red) and
\unit[1.2]{\um} (blue). The colored stripes highlight regions in
field where the resistance in the \unit[1.2]{\um} and
\unit[0.8]{\um} \qpc s forms a plateau-like feature near
$\nuqpc=\fhf$ with $\nubulk=3$. The applied gate voltages \Vg\ are
-2.2, -2.0 and \unit[-1.9]{V} for the 1.2, 0.8 and \unit[0.5]{\um}
\qpc s and the ac lock-in excitation is \unit[0.86]{nA}. All data is
for $T\sim\unit[8]{mK}$.}
\end{figure}

The edge state interpretation for \Rxy, \Rxx, \Rd, and \Rl\ has been
extended to the \fqhe\
\cite{Beenakker1989,MacDonald1990,BandVRd,Chang1989,kouwenhoven1990,WenPRL1990,WangGoldman1991,Wurtz2004}.
Within this generalized picture, a quantized plateau in
$\Rxy\sim\runit (1/\nubulk)$ corresponds to the quantum Hall state
at filling fraction $\nubulk$, and a plateau in
$\Rd\sim\runit(1/\nuqpc)$ indicates that an incompressible quantum
Hall state has formed in the vicinity of the \qpc\ with effective
filling fraction \nuqpc. We associate deviations from precisely
quantized values with tunneling, which we study below as a function
of temperature and bias.

To simplify the study of quantum states in the vicinity of the \qpc,
the perpendicular magnetic field ($B$) and gate voltage of the \qpc\
(\Vg) are tuned such that \nubulk\ is fixed at an integer quantum
Hall effect (\iqhe) plateau whenever \nuqpc\ is at a value of
interest. With $\Rxx \sim 0$ and \Rxy\ quantized to an \iqhe\
plateau, features in \Rd\ and \Rl\ measurements can be attributed to
the \qpc\ region and not the bulk.

\begin{figure}[t]
\center
\includegraphics{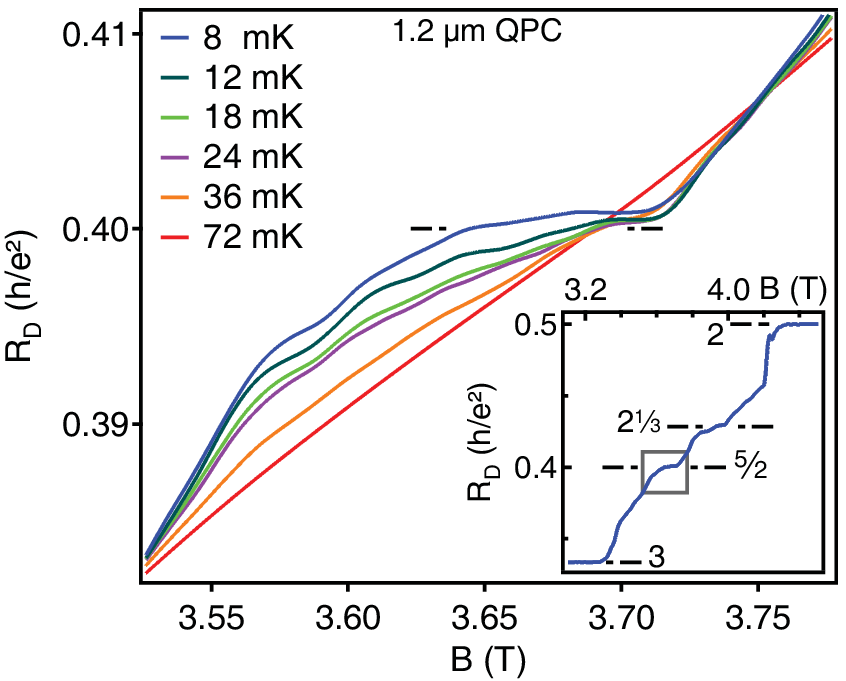}
\caption{{\label{Temp}} Temperature dependence of the \fhf\ state in
the \unit[1.2]{\um} \qpc.  The inset shows an expanded range of the
\unit[8]{mK} trace with the grey box indicating the range of the
data in the main panel. All traces are measured with
\Vg=\unit[-2.7]{V} and an ac lock-in excitation of \unit[0.86]{nA}.
$\nubulk=3$ for the entire $B$ range of the main panel, but not the
full range of the inset.}
\end{figure}

Previously, \qpc s have been used to selectively transmit integer
\cite{vanwees1988,alphenaar1990} and fractional edge channels
\cite{kouwenhoven1990,alphenaar1992}, and to study inter-edge
tunneling between fractional edge channels, including in the regime
where the bulk is intentionally set to an \iqhe\ plateau
\cite{roddaro:156804,lal2006}. Comparisons with these results are
discussed below. \qpc s have also been employed in studies of noise
\cite{Saminadayar1997,de-Picciotto} and (along with etched trenches)
interference of quasiparticles \cite{camino:246802} in the \fqhe\
regime. In all of these studies $\nu<2$, where the \fqhe\ gaps are
typically much larger than those with $\nu>2$.

\begin{figure}[t]
\center
\includegraphics{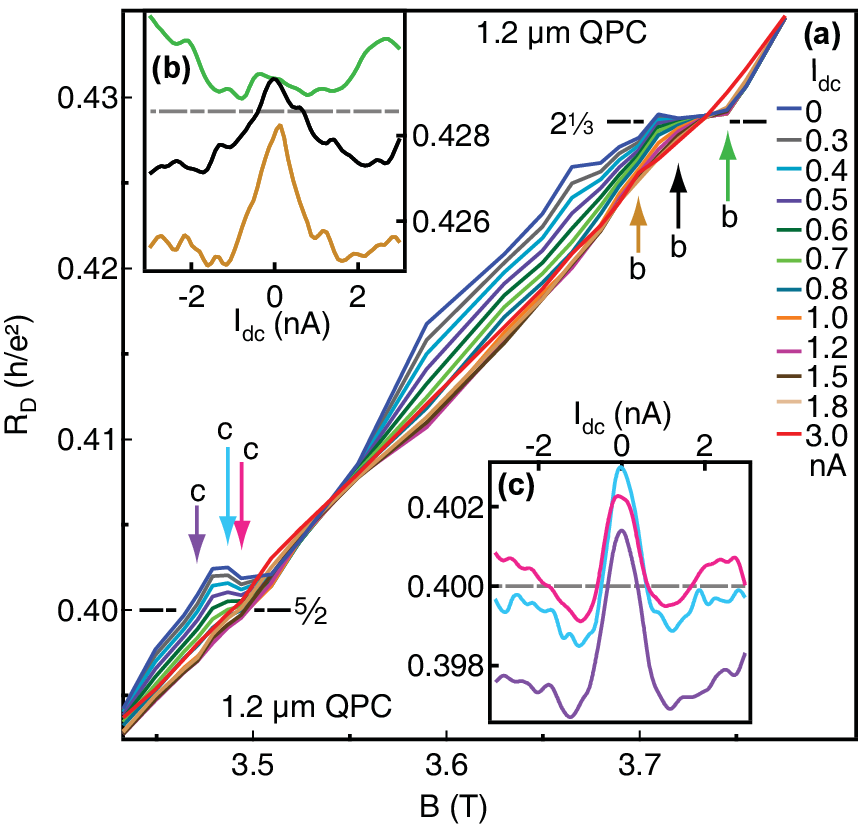}
\caption{{\label{Idc}} Dependence upon dc current bias of the \fhf\
and $2\nicefrac{1}{3}$ states in the \unit[1.2]{\um} \qpc. The main
panel (a) shows the \Rd\ data as a function of magnetic field; each
trace represents a different \Idc from $\unit[0]{nA}$ to
$\unit[3]{nA}$. \Rd\ as a function of \Idc\ for selected magnetic
fields (indicated by the color-coded arrows) are shown in (b) and
(c). The dotted grey lines in the insets indicate resistance values
of $\nicefrac{3}{7} \runit$ (b) and $\nicefrac{2}{5} \runit$ (c).
All traces are measured at $T\sim\unit[8]{mK}$ with
$\Vg=\unit[-2.4]{V}$ and an ac lock-in excitation of \unit[0.2]{nA}.
$\nubulk=3$ for all fields shown in this figure.}
\end{figure}

The sample is a GaAs/AlGaAs heterostructure grown in the [001]
direction with electron gas layer \unit[200]{nm} below the surface,
with Si $\delta$-doping layers \unit[100]{nm} and \unit[300]{nm}
below the surface.  A \unit[150]{\um}-wide Hall bar is patterned
using photolithography and a H$_2$O:H$_2$SO$_4$:H$_2$O$_2$ (240:8:1)
wet-etch, followed by thermally evaporated Cr/Au
(\unit[5]{nm}/\unit[15]{nm}) top-gates patterned using electron-beam
lithography (see Fig.~\ref{Device}). The gates form \qpc s with
lithographic separation between gates of 0.5, 0.8 and
\unit[1.2]{\um}. Depleting the electron gas beneath only one side of
a \qpc\ has no effect on transport measurements. Gate voltages were
restricted to the range \unit[-1.9]{V} (depletion) to \unit[-3]{V}
and were allowed to stabilize for several hours at each setpoint
before measuring; beyond \unit[-3]{V} the conductance was typically
hysteretic as a function of gate voltage. Measurements are performed
in a dilution refrigerator with base temperature \unit[6]{mK} using
standard four-wire lock-in techniques, with an ac current-bias
excitation (\Iac) ranging from \unit[0.2]{nA} to \unit[0.86]{nA},
and a dc current-bias ranging from $0$ to \unit[20]{nA}.  The
differential resistances ($dV/d\Iac$) are measured in four places,
as shown in Fig.~\ref{Device}. All quoted temperatures are measured
using a RuO$_2$ resistor mounted on the mixing chamber. The bulk
mobility of the device measured at base temperature is
\unit[2000]{m$^2$/Vs} and the electron density is
\unit[2.6$\times$10$^{15}$]{m$^{-2}$}.

Bulk \Rxx\ and \Rxy\ measurements for the filling fraction range
$\nubulk=3$ to 2, measured in the vicinity of the \unit[1.2]{\um}
\qpc\ before the gates are energized, are shown in Fig.~\ref{Bulk}.
\Rxx\ and \Rxy\ are also  measured in a region of the Hall bar
without gates, and found to be virtually indistinguishable, showing
that the surface gates do not significantly affect the \twodeg.
\Rxx\ and \Rxy\ in an un-gated region show no changes caused by
energizing gates.

As temperature is increased, \Rxy\ near $\nubulk=\fhf$ evolves from
a well-defined plateau at \Rxy=\unit[$0.4 \pm 0.0002$]{\runit} to a
line consistent with the classical Hall effect for a material with
this density. There is a stationary point in the middle of the
plateau where \Rxy\ is very close to \unit[0.4]{\runit}, consistent
with scaling seen in other quantum Hall transitions
\cite{dassarmain}. Activation energies $\Delta$ for the three
fractional states $\nubulk=\fhf$, 2$\nicefrac{1}{3}$ and
$2\nicefrac{2}{3}$ are extracted from the linear portion of the data
in a plot of $\ln(\Rxx)$ vs $1/T$ (using the minimum \Rxx\ for each
\fqhe\ state, and $\Rxx \propto e^{-\Delta/{2T}}$), giving
$\Delta_{\dipL}\sim\unit[60]{mK}$, $\Delta_{\fhf}\sim\unit[130]{mK}$
and $\Delta_{\dipR}\sim\unit[110]{mK}$, consistent with previous
measured values \cite{Willett1987,eisenstein2002,xia2004}.

We now focus on measurements with one \qpc\ formed, as shown in
Fig.~\ref{Device}. Low-field \Rd\ and \Rl\ data from the
\unit[1.2]{\um} \qpc\ along with concurrently measured \Rxy\ and
\Rxx\ show regions where one \iqhe\ state forms in the bulk with a
lower \iqhe\ state in the \qpc\ (see Fig.~\ref{LowB}a and
Fig.~\ref{LowB}b). Figure~\ref{LowB}a also shows the appearance of a
plateau-like feature in the \qpc\ between $\nuqpc=5$ and $\nuqpc=4$
in both the \unit[1.2]{\um} and \unit[0.8]{\um} \qpc s which remains
unexplained.  At higher magnetic fields (Fig.~\ref{LowB}c and
Fig.~\ref{LowB}d), \Rd\ and \Rl\ show \fqhe\ plateaus while the bulk
is quantized at the \iqhe\ value $\nubulk=2$.

We now concentrate on the range $\nuqpc=3$ to $\nuqpc=2$ with
$\nubulk=3$ (Fig.~\ref{ManyQPC}). Plateau-like structure near
$\nuqpc=\fhf$ is evident in the \unit[1.2]{\um} and \unit[0.8]{\um}
\qpc s, but is not seen in the \unit[0.5]{\um} \qpc. Near
$\nuqpc=\dipR$ we also see plateau-like behavior in the
\unit[1.2]{\um} \qpc, and somewhat less well developed plateaus in
in the \unit[0.8]{\um} \qpc\ (although \nubulk\ is not on a plateau
when $\nuqpc \sim \dipR$), but again these features are suppressed
in the \unit[0.5]{\um} \qpc. We do not observe any plateaus near
$\nuqpc=\dipL$ in any of the \qpc s. The reentrant integer quantum
Hall effect features, which are clearly visible in the bulk, do not
survive at all in the \qpc s.

We interpret the plateau-like features in the two larger \qpc s as
indicating that the incompressible states at $\nuqpc=\fhf$ and
$\nuqpc=\dipR$ are not destroyed by the confinement. The linear,
plateau-less behavior in the \unit[0.5]{\um} \qpc\ is reminiscent of
a classical Hall line, suggesting that no incompressible states
survive in this \qpc.

Temperature dependence for a representative \Vg\ setting of the
\unit[1.2]{\um} \qpc\ is shown in Fig.~\ref{Temp}. Below
\unit[30]{mK}, a distinct plateau-like feature is evident. This
plateau disappears between 30 to \unit[70]{mK}, consistent with the
disappearance of the plateaus in the bulk. However, unlike the bulk,
where the \fhf\ plateau disappears symmetrically around a stationary
point at $\Rxy=$~\unit[0.4]{\runit} as temperature increases, in the
\qpc\ there is an additional resistance: \Rd\ exceeds the quantized
value of \unit[0.4]{\runit} by
$\unit[26]{\Omega}\pm\unit[5]{\Omega}$. We also note that the the
extra resistance on the plateau decreases as the temperature
increases, behavior consistently observed in both the
\unit[0.8]{\um} and \unit[1.2]{\um} \qpc s. We interpret this as
indicating that the temperature dependence comes not only from the
thermal excitation of quasiparticles, but also from the temperature
dependence of their backscattering.

The dependence of the differential resistance on dc source-drain
bias \Idc\ (Fig.~\ref{Idc}) provides additional insight into this
excess resistance. At base temperature, the resistances \Rd\ vs
\Idc\ near $\nuqpc=\fhf$ and $\nuqpc=\dipR$ in the \unit[1.2]{\um}
(Fig.~\ref{Idc}c) and \unit[0.8]{\um} (not shown) \qpc s show
pronounced peaks at $\Idc=0$, a dip at intermediate values, and
saturation to a constant value at high currents. In these \qpc s,
the \Idc\ behavior near $\nuqpc=\dipL$ (not shown) is inverted, with
a pronounced dip at $\Idc=0$ a peak at intermediate values, and
high-current saturation. In the \unit[0.5]{\um} \qpc\ the \Idc\
traces are flat for all filling fractions between $\nuqpc=3$ and
$\nuqpc=2$. All the traces in Fig.~\ref{Idc} are measured with an
ac-lock-in excitation $\Iac=\unit[0.2]{nA}$, (while the data in all
other figures have been measured with $\Iac=\unit[0.86]{nA}$).

Fig.~\ref{Idc} provides a key point of comparison to previous
experimental and theoretical work on the \fqhe. In a recent
experiment \cite{roddaro:156804}, a \qpc\ is used to measure
tunneling differential resistance characteristics (\Idc\ curves) for
$\nuqpc<1$ while $\nubulk$ is fixed on an \iqhe\ plateau. Our \Idc\
data for $2<\nuqpc<3$ and $\nubulk=3$, with a distinct peak at zero
bias and dips at intermediate biases, resembles the \Idc\ curves in
that work. In Ref.~\citealp{roddaro:156804} it is convincingly
argued that the \Idc\ curves are a signature of quasiparticle
tunneling between the \fqhe\ edge states, based on quantitative
comparison to applicable theory. That theory states that the
characteristic for tunneling between \fqhe\ edge states
\cite{WenPRL1990,Moon1993,FendleyPRB1995} is expected to have a peak
at zero bias and a minimum at intermediate biases, whereas tunneling
between \iqhe\ edge channels is expected to yield a flat (ohmic)
curve. The data we present for $\nuqpc=\fhf$, both the temperature
dependence and the \Idc\ curves, are consistent with the formation
of a \fqhe\ state with tunneling-related backscattering.

We interpret that a mechanism for the deviation of \Rd\ from
\unit[0.4]{\runit} near \fhf\ and \dipR, as well as the peak-and-dip
behavior of the \Idc\ data, could be tunneling between edge channels
on opposite sides of Hall bar in the vicinity of the \qpc. We do not
believe the data can be explained by transport via thermally excited
particles through the (small) bulk region of the \qpc, since this
process would be expected to have the opposite temperature
dependence.

We gratefully acknowledge helpful discussions with M.\ Fisher, B.\
Halperin, A.\ Johnson, E.\ Kim,  B.\ Rosenow, A.\ Stern, X.-G.\ Wen
and A.\ Yacoby. Research supported in part by Microsoft Corporation
Project Q, and HCRP at Harvard University, and ARO
(W911NF-05-1-0062), the NSEC program of the NSF (PHY-0117795) and
NSF (DMR-0353209) at MIT.

\end{document}